\documentclass[nofootinbib,nobibnotes,groupaddress,preprintnumbers,aps,pre,twocolumn]{revtex4}
\usepackage{tipa}
\usepackage{color}
\usepackage{bm,epsfig,mathrsfs,amsmath,amssymb,graphicx,subfigure,overpic}
\usepackage{graphicx}
\usepackage{dcolumn}
\usepackage{bm,ulem}

\usepackage{chngcntr}
\usepackage{float} 
\usepackage{booktabs}%
\newcommand{\PreserveBackslash}[1]{\let\temp=\\#1\let\\=\temp}
\newcolumntype{C}[1]{>{\PreserveBackslash\centering}p{#1}}
\newcolumntype{R}[1]{>{\PreserveBackslash\raggedleft}p{#1}}
\newcolumntype{L}[1]{>{\PreserveBackslash\raggedright}p{#1}}
\usepackage[colorlinks,citecolor=blue,linkcolor=red]{hyperref}
\usepackage[toc,title,titletoc,header]{appendix}

\date{\today}
\begin{document}
\title{Information-Assisted Carnot Engine Surpasses Standard Thermodynamic Bounds}
\author{Yang Xiao$^{1}$}
\author{Qian Zeng$^2$}
\author{Jin Wang$^{3}$}\email{jin.wang.1@stonybrook.edu}
\affiliation{ $^1\,$College of Physics, Jilin University, Changchun 130022, China\\ $^2\,$   State Key Laboratory of Electroanalytical Chemistry, Changchun Institute of Applied Chemistry, Changchun, Jilin 130022, China\\ $^3\,$ Department of Chemistry and Department of Physics and Astronomy, State University of New York at Stony Brook, Stony Brook, New York 11794, USA}

\begin{abstract} 
Information can improve heat engine performance, but the underlying principles are still not so clear. Here we introduce a Carnot information machine (CIE) and obtain a quantitative relationship between the engine  performance and information. 
We demonstrate that the presence of information changes allows the CIE to operate as a heat engine in the regime where the standard Carnot cycle is prohibited, ensures that the efficiency of the CIE is greater than or equal to the standard Carnot efficiency, and significantly enables it to achieve 100\% efficiency with positive work extraction for arbitrary two-level systems. We explicitly demonstrate these features using a spin-1/2 system and propose an experimental implementation scheme based on a trapped $^{40}\mathrm{Ca}^+$ ion.
\end{abstract}
\maketitle
\date{\today}

\textit{Introduction}.$-$The information engine, which utilizes information to extract work from  heat reservoirs \cite{Szilard1929,Sagawa10,JM19,SY16,LB19,Chen21,Ding18,JY96,JJ13,CE17,CE18,Lutz23,SS20,KC17,JV15,LB21,Jar23,Jar12,TK21,TK23}, was first conceptualized by Szilard \cite{Szilard1929} as a solution to Maxwell's demon paradox \cite{Maxwell1871,Vedral09}. Far from being merely a theoretical construct, such information engines represent fundamental mechanisms widespread in nature \cite{EM21,Wen18,Sc24,SF23,VS07}. Among these, measurement-feedback engines, also known as Maxwell's demon-like mechanisms, are particularly significant. In these engines, the demon controls the working system's evolution based on information stored in its memory \cite{Sagawa10,JJ13,CE17,CE18,Lutz23,SS20,KC17,JV15,LB21}, a process that has profoundly influenced the development of modern thermodynamics \cite{Sagawa10,sagawa12,Sagawa08,GP20,TV20,Zeng21}. These naturally occurring mechanisms have been extensively studied and experimentally realized across various physical systems, 
 including spin-boson \cite{SS20}, Brownian motion \cite{BJ08, Sag10, GP18, YJ14}, electronic  \cite{JV15, JV14, Sag14, KC17}, photonic \cite{MD16}, and superconducting circuits system \cite{NC17}.

Conventional heat engines operating between thermal reservoirs at inverse temperatures $\beta_\mathrm{h}$ and $\beta_\mathrm{c}$ ($\beta_\mathrm{c}>\beta_\mathrm{h}$) convert thermal energy  into mechanical work 
and its efficiency is  fundamentally limited by the Carnot efficiency $\eta_\mathrm{C}=1-\beta_\mathrm{h}/\beta_\mathrm{c}$ \cite{Car90,Cal85}.
Although information-assisted cycles can enable efficiencies beyond this limit \cite{Sagawa08,SR16,Jar13}, 
the mechanism by which information achieves this remains elusive. The critical challenge lies in the lack of a quantitative relationship between information and heat engine performance (output work and efficiency).  The key to achieving this hinges on simultaneously regulating heat engine performance and mutual information.

To address these crucial challenges, we propose a Carnot information engine (CIE), incorporating an imperfect classical measurement and feedback control into the standard Carnot cycle. 
Our analytical framework establishes explicit relations between mutual information changes and  work output as well as efficiency. Specifically, our theoretical analysis demonstrates that: (i) information changes enable the CIE to operate as a heat engine in regimes where the standard Carnot cycle is forbidden, which is achieved through entropy flow compensation via measurement and feedback; (ii) information changes make the efficiency of the CIE be greater than or equal to the standard Carnot efficiency; and (iii) most significantly, the CIE can achieve 100\% efficiency while producing positive work, breaking the efficiency limits of conventional heat engines. To validate these theoretical findings, we illustrate our results with a detailed example of a spin-1/2 CIE and present an experimental implementation scheme based on a trapped $^{40}\mathrm{Ca}^+$ ion.

\begin{figure*}
\centering
\begin{overpic}[width=16cm]{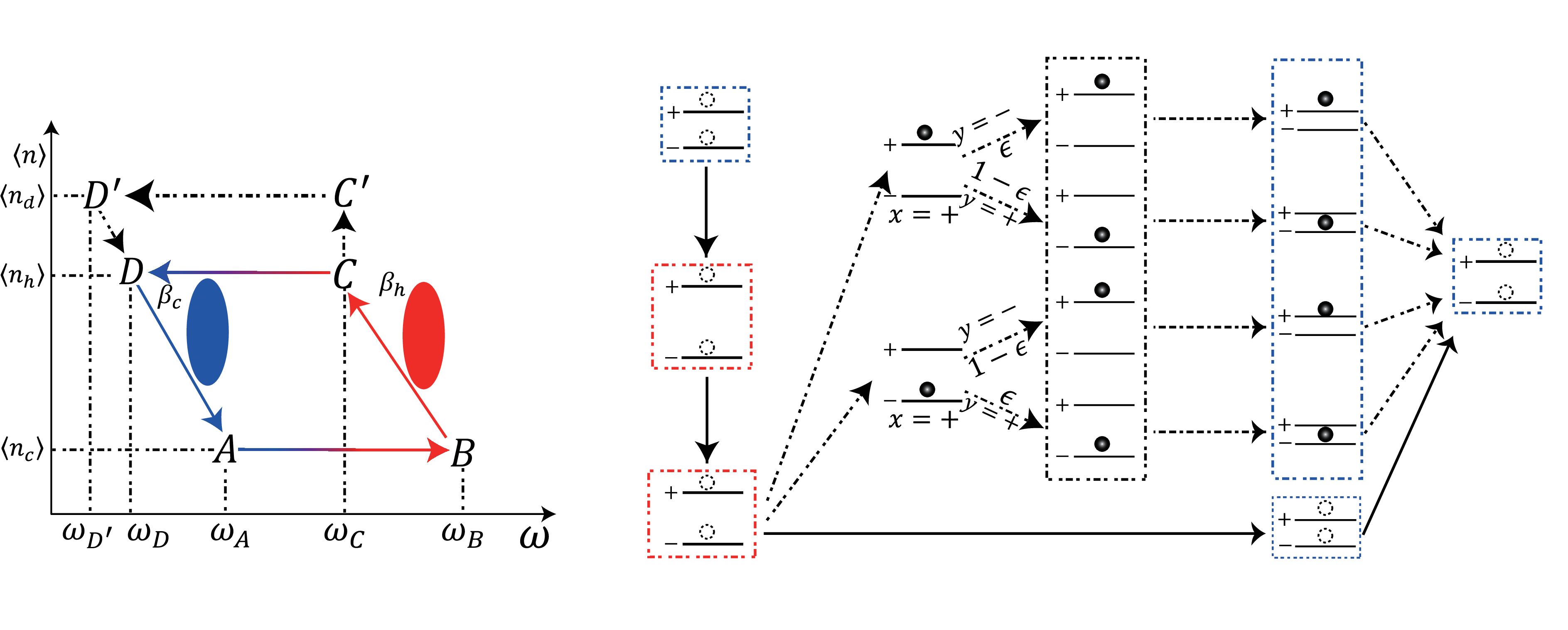}
\put(37.5,24){\scalebox{1}{\rotatebox{90}{adiabatic}}}
\put(39,23){\scalebox{1}{\rotatebox{90}{compression}}}
\put(37.5,9.7){\scalebox{1}{\rotatebox{90}{isothermal}}}
\put(39,10){\scalebox{1}{\rotatebox{90}{expansion}}}
\put(60,4){\scalebox{1}{adiabatic expansion}}
\put(89,9){\scalebox{1}{\rotatebox{66}{expansion}}}
\put(90.5,8){\scalebox{1}{\rotatebox{66}{isothermal}}}
\put(57,39.5){\scalebox{1}{measurement and control}}
\put(0,34.5){\scalebox{1.2}{(a)}}
\put(38,34.5){\scalebox{1.2}{(b)}}
\put(44,35.2){\scalebox{1.2}{$A$}}
\put(38,20){\scalebox{1.2}{$B$}}
\put(38,6.5){\scalebox{1.2}{$C$}}
\put(82.7,2.5){\scalebox{1.2}{$D$}}
\put(69.3,37){\scalebox{1.2}{$C^{'}$}}
\put(82.7,37){\scalebox{1.2}{$D^{'}$}}
\put(95,26){\scalebox{1.2}{$A$}}
\end{overpic}
    \caption{Schematic of the CIE. (a) The average population-frequency diagram for  an arbitrary two-level work system. The standard Carnot cycle consists of four consecutive strokes: adiabatic compression  $A\rightarrow B$, isothermal expansion $B\rightarrow C$, adiabatic expansion $C\rightarrow D$,  isothermal compression $D\rightarrow A$. The CIE contains five consecutive strokes: adiabatic compression $A\rightarrow B$, isothermal expansion $B\rightarrow C$, measurement and control $C\rightarrow C^{'}$, adiabatic expansion $C^{'}\rightarrow D^{'}$, and 
 isothermal compression $D^{'}\rightarrow A$. 
 (b) Explicit realization of the spin-1/2 CIE. The solid (dashed) arrows represent the standard  Carnot cycle (CIE) and the solid (dashed) balls represent the system state being known (unknown)}
    \label{model}
\end{figure*}
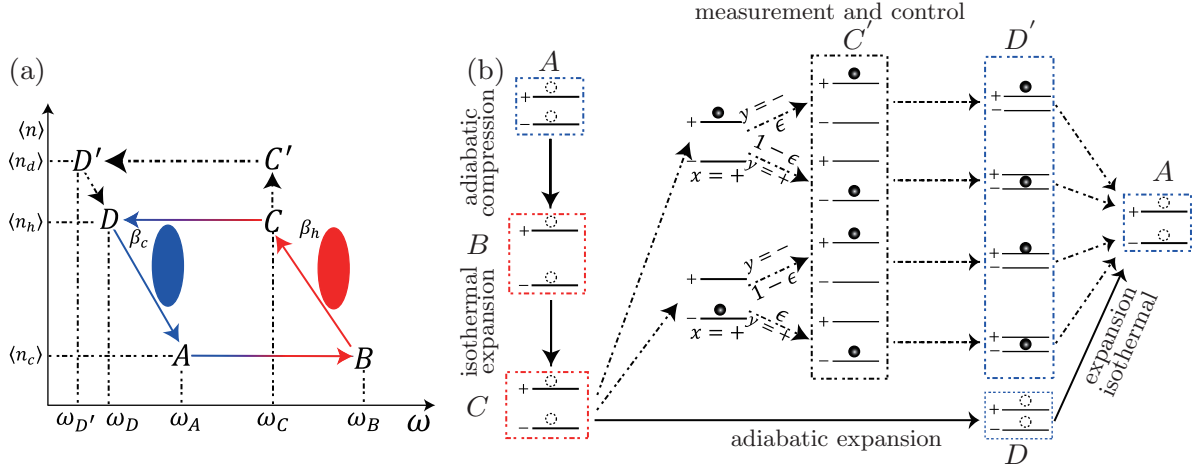

\textit{Carnot information  engine cycle}.$-$  In this heat engine, the working system's energy levels are characterized by the populations in excited and ground states, denoted by $n_e$, and $n_g$,  respectively, with corresponding energy levels defined as $\varepsilon^{e}=\omega n_e$, $\varepsilon^{g}=\omega n_g$, where $\omega$ represents the system's frequency.  Let $x$ denote the system state. When the system reaches thermal equilibrium with a heat reservoir at inverse temperature $\beta$, the occupation probabilities of the excited and ground states are determined as $P^e=P(x=e)=\exp(-\beta \varepsilon^e)/Z
$ and $P^g=P(x=g)=\exp(-\beta \varepsilon^g)/Z$ respectively, where  $Z=\exp(-\beta \varepsilon^g)+\exp(-\beta \varepsilon^e)$ is the partition function.

The CIE operates through five distinct strokes, as illustrated in Fig. \ref{model}(a):  two standard Carnot cycle strokes (adiabatic compression stroke $A\rightarrow B$ and isothermal expansion stroke $B\rightarrow C$), followed by three control strokes $C\rightarrow C^{'} \rightarrow D^{'}\rightarrow A$. 
During the adiabatic compression stroke $A\rightarrow B$, the system undergoes a compression from  $\omega_A$  to $\omega_B$  while maintaining constant population numbers. In the subsequent isothermal expansion stroke $B\rightarrow C$, the system interacts with the hot reservoir at inverse temperature $\beta_\mathrm{h}$,  the system frequency decreases reversibly from $\omega_B$ to $\omega_C$. This process results in heat absorption by the system be proportional to the temperature times the entropy change \cite{HT07}
\begin{equation} 
Q_\mathrm{h}
    =\beta_\mathrm{h}^{-1}(
    S_C-S_B),\label{Qh}
\end{equation}
 where,  $S_B=S_A=S_\mathrm{c}=-P_A^e\ln P_A^e -P_A^g\ln P_A^g$ and 
  $S_C=S_\mathrm{h}=-P_C^e\ln P_C^e -P_C^g\ln P_C^g$  represent  the  entropy of the system at points $B$ and $C$. 
  
  The  remaining three strokes are controlled by the demon through measurement and feedback. At point $C$, the demon  performs a measurement on the system state $x_C$ ($x_C=e,g$) with a measurement
error characterized by  $p(y\neq x_C|x_C)=\epsilon$. Here, $y=e,g$ denotes the measurement outcome. Based on this measurement, the demon implements a state-dependent control protocol. 
When $y=g$, the control protocol consists of three consecutive strokes: (a) $C\rightarrow C^{'}$:  The system state is maintained unchanged. (b) $C^{'}\rightarrow D^{'}$: The system undergoes an adiabatic expansion while keeps constant state. (c) $D^{'}\rightarrow A$: The system interacts with the cold reservoir at
inverse temperature $\beta_\mathrm{c}$ and returns to its initial state.
Alternatively, when $y = e$, the demon first flips the system's energy level during $C\rightarrow C^{'}$, then performs the same subsequent control sequence as in the $y = g$ case \cite{sup1}. The detailed dynamics of these control strokes are explained in the following section.

The first control stroke $C\rightarrow C^{'}$:  Following the measurement, the system becomes correlated with the demon.  The  mutual information between  them is given by (see  Appendix \ref{inf})
\begin{equation}\label{IC}
 I_C=S_Y-S_\epsilon,
\end{equation}
 where $S_Y=-P_y^g\ln P_y^g-P_y^e\ln P_y^e$ represents the demon's entropy and  $S_\epsilon=-(1-\epsilon)\ln(1-\epsilon)-\epsilon\ln\epsilon$ denotes the entropy of the measurement error. Here $P^e_y=P(y=e)=P_C^e(1-\epsilon)+P_C^g\epsilon$, and $P^g_y=P(y=g)=P_C^e\epsilon+P_C^g(1-\epsilon)$ represent the probabilities of measurement outcomes $y=e$ and $y=g$, respectively. 
 
 Then  the demon utilizes the obtained information $ I_C$ to perform work by flipping the  system's energy level while  fixing the system  frequency at $\omega_C$. This operation transforms the system's probability distribution to
 $P^e_{C^{'}}=
   P^e_CP(y=g|x_C=e)
   +P^g_CP(y=e|x_C=g)
   =\epsilon $ and $ P^g_{C^{'}}=P^e_CP(y=e|x_C=e)
  +P^g_CP(y=g|x_C=g)=1-\epsilon,$  where the system's entropy becomes to the measurement error entropy $S_\epsilon$.
Through   the  detailed balance condition $\exp{[\beta_\mathrm{d}(\varepsilon^e_C-\varepsilon^g_C)]}=P^g_{C^{'}}/P^e_{C^{'}}$,  
the effective inverse temperature $\beta_\mathrm{d}$ after this control stroke can be determined.
Based  on the distribution $P_{C^{'}}^e$ and  $P_{C^{'}}^g$,  the work exerted by the demon on the system  can be calculated as (see Appendix \ref{inf})
\begin{eqnarray}\label{Wd}
    W_\mathrm{d}
&=&\mathop{\sum}\limits_{x_C,x_{C^{'}},y} P(x_C\rightarrow x_{C^{'}}|y)P(y)(\varepsilon_C^{x_{C^{'}}}-\varepsilon_C^{x_C})\nonumber\\
    &=&\omega_C(\langle n_\mathrm{d}\rangle- \langle n_\mathrm{h}\rangle),
\end{eqnarray}
where $P(x_C\rightarrow x_{C^{'}}|y)$ is the condition probability for the 
system's evolutionary trajectory from $x_C$ to $x_{C^{'}}$ based on   the measurement outcome $y$. Due to this control stroke is nondegenerate, this probability  is equal to the condition probability $P(x_C|y)$.
Here $ \langle n_\mathrm{h}\rangle=P_C^en_e+P_C^gn_g$ and  $\langle n_\mathrm{d}\rangle=P^e_{C^{'}}n_e+P^g_{C^{'}}n_g$  represent the average populations of the system at points $C$ and $C^{'}$, respectively.   After being utilized by the demon, the remaining mutual information at point $C^{'}$ can be expressed as (see Appendix \ref{inf})
\begin{equation}\label{IC1}
 I_{C^{'}}=S_Y-S_\mathrm{h}.
\end{equation}
Then, the information changes during this process are given by $ \Delta I = I_{C^{'}}- I_C=S_\epsilon-S_\mathrm{h}$.  This quantity characterizes how the demon modulates the system entropy
 through its imperfect measurement and control operations.

The second control stroke $C^{'}\rightarrow D^{'}$: during this stroke, the demon fixes the system state at the final state of the conditional trajectory $x_C\rightarrow x_{C^{'}}|y$. Simultaneously, the demon guides the system to  expand adiabatically from $\omega_C$ to $\omega_{D^{'}}$. This operation indicates the conditional probability $P(x_C\rightarrow x_{D^{'}}|y)=P(x_C\rightarrow x_{C^{'}}|y)=P(x_{C^{'}}|y)$. Consequently, the system probability distribution at the point $D^{'}$ is the same as at point $C^{'}$, i.e., $P_{x_{D^{'}}=x_{C^{'}}}=P_{x_{C^{'}}}$. At the end of this stroke, the system reaches thermal  equilibrium with the cold reservoir.  Through  the detailed balance condition $\exp[\beta_\mathrm{c}(\varepsilon_{D^{'}}^e-\varepsilon_{D^{'}}^g)]=P^g_{D^{'}}/P^e_{D^{'}}$, the system  frequency $\omega_D^{'}$ modulated by the demon can be determined.

The third control stoke $D^{'}\rightarrow A$: The demon couples the system to the cold reservoir with the inverse  temperature $\beta_\mathrm{c}$ while modulating the system frequency from $\omega_D^{'}$ to its initial quantity $\omega_A$.  Throughout this process, the system maintains equilibrium with the cold reservoir, causing the system's distribution to become independent of the demon's control.  
The heat absorbed from the cold reservoir  can be expressed through the entropy flow from the system to the cold reservoir:
\cite{Zeng21,TC00} 
\begin{eqnarray}\label{Qc1}
  Q_\mathrm{c}^\mathrm{d} &=&\mathop{\sum}\limits_{x_{D^{'}},x_A,y}(\beta_\mathrm{c})^{-1}P({x}_{D^{'}}\rightarrow x_A|x_{D^{'}},y)\nonumber\\
&\times&P(x_{D^{'}}|y)P(y)\ln\frac{P(x_A\rightarrow{x}_{D^{'}}|x_A,y)}{P({x}_{D^{'}}\rightarrow x_A|x_{D^{'}},y)}\nonumber\\
&=&\beta_\mathrm{c}^{-1}(S_\mathrm{c}-S_\epsilon)
=Q_\mathrm{c}-\beta_\mathrm{c}^{-1} \Delta I ,
\end{eqnarray} 
where $Q_\mathrm{c}=\beta_\mathrm{c}^{-1}(S_\mathrm{c}-S_\mathrm{h})$  represents the heat absorbed from the cold reservoir during the isothermal compression stroke  $D\rightarrow A$ of the standard Carnot cycle as shown in Fig. \ref{model}(a). Note that,  $P(x_{D^{'}}\rightarrow x_A|x_{D^{'}},y)$ quantifies  the condition probability for the 
system's evolutionary trajectory from $x_{D^{'}}$ to $x_{A}$, based on the measurement outcome $y$.  Because the system and the demon's memory become uncorrelated during this stroke,   this probability is equal to the initial probability $P(x_A)$.  Similarly, the probability for time-reversal trajectory  $P(x_A\rightarrow x_{D^{'}} |x_A,y)$ is  equal to  $P(x_{D^{'}})$.
The last 
equality of Eq. (\ref{Qc1}) allows 
us to decompose the isothermal compression stroke  $D^{'}\rightarrow A$ into two distinct subsections: the demon-induced compression  $D^{'}\rightarrow D$, and the standard Carnot compression $D\rightarrow A$. In the first subsection, the system releases a net heat $\Delta I/\beta_\mathrm{c}$  into the cold reservoir, which compensates for the demon-induced entropy change during the first control stroke. Thus, the CIE can be regarded as a combination of the standard Carnot cycle $A\rightarrow B\rightarrow C\rightarrow D\rightarrow A$ and the information-driven cycle $C\rightarrow C^{'}\rightarrow D^{'}\rightarrow D\rightarrow C$. 
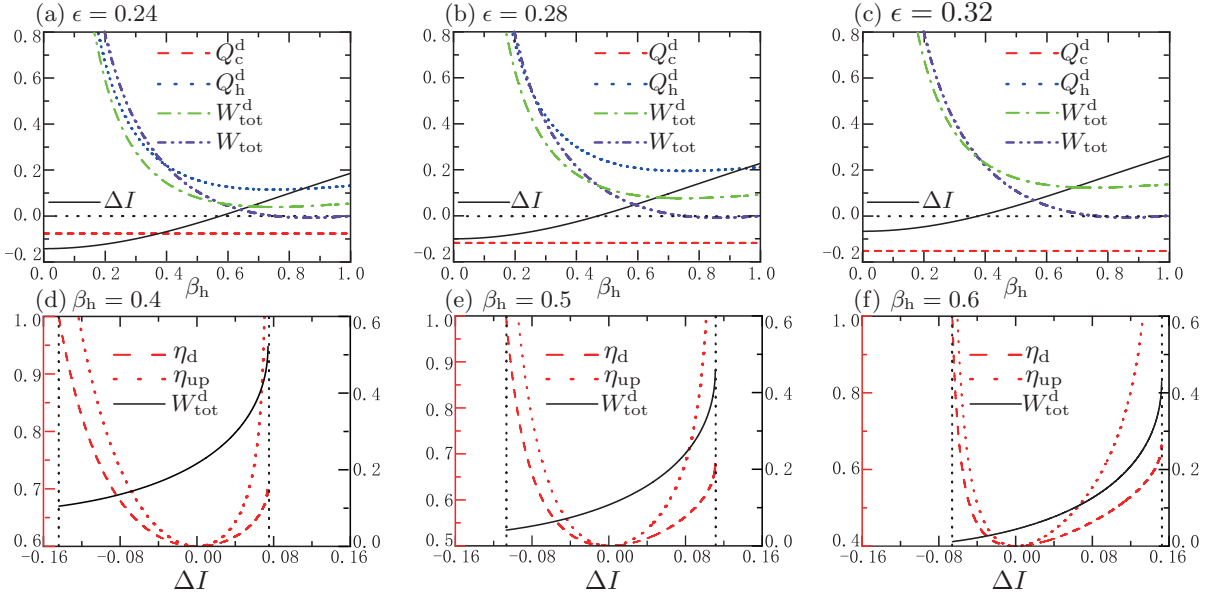
\begin{figure*}
\centering
\begin{overpic}[width=16cm]{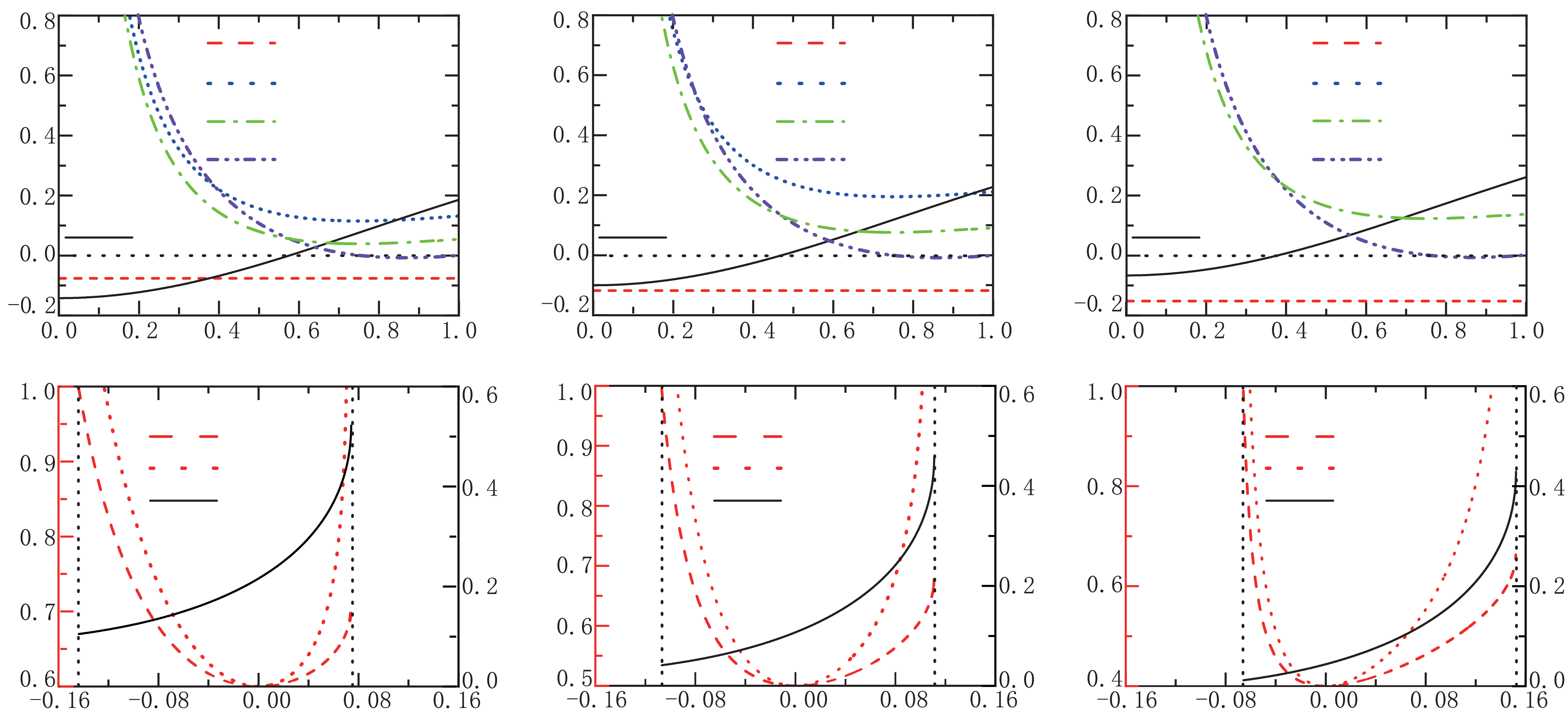}
\put(3,45.8){\scalebox{1}{(a)}}
\put(6.1,45.8){\scalebox{1}{$\epsilon=0.24$}}
\put(37,45.8){\scalebox{1}{(b)}}
\put(40.2,45.8){\scalebox{1}{$\epsilon=0.28$}}
\put(71,45.8){\scalebox{1}{(c)}}
\put(74.1,45.8){\scalebox{1.2}{$\epsilon=0.32$}}

\put(15.5,23){\scalebox{1}{$\beta_\mathrm{h}$}}
\put(49.7,23){\scalebox{1}{$\beta_\mathrm{h}$}}
\put(83.5,23){\scalebox{1}{$\beta_\mathrm{h}$}}

\put(18,35.1){\scalebox{1}{$ W_\mathrm{tot}$}}
\put(54,35.1){\scalebox{1}{$W_\mathrm{tot}$}}
\put(88.2,35.1){\scalebox{1}{$W_\mathrm{tot}$}}

\put(18,37.5){\scalebox{1}{$W_\mathrm{tot}^\mathrm{d}$}}
\put(54,37.5){\scalebox{1}{$W_\mathrm{tot}^\mathrm{d}$}}
\put(88.2,37.5){\scalebox{1}{$W_\mathrm{tot}^\mathrm{d}$}}

\put(18,40){\scalebox{1}{$ Q_\mathrm{h}^\mathrm{d}$}}
\put(54,40){\scalebox{1}{$ Q_\mathrm{h}^\mathrm{d}$}}
\put(88.2,40){\scalebox{1}{$ Q_\mathrm{h}^\mathrm{d}$}}

\put(18,42.5){\scalebox{1}{$Q_\mathrm{c}^\mathrm{d}$}}
\put(54,42.5){\scalebox{1}{$Q_\mathrm{c}^\mathrm{d}$}}
\put(88.2,42.5){\scalebox{1}{$ Q_\mathrm{c}^\mathrm{d}$}}

\put(8.7,30.2){\scalebox{1}{$\Delta {I}$}}
\put(42.5,30.2){\scalebox{1}{$\Delta {I}$}}
\put(76.7,30.2){\scalebox{1}{$\Delta {I}$}}

\put(3,22){\scalebox{1}{(d)}}
\put(6.1,22){\scalebox{1}{$\beta_\mathrm{h}=0.4$}}
\put(37,22){\scalebox{1}{(e)}}
\put(40.2,22){\scalebox{1}{$\beta_\mathrm{h}=0.5$}}
\put(71,22){\scalebox{1}{(f)}}
\put(74.1,22){\scalebox{1}{$\beta_\mathrm{h}=0.6$}}

\put(14.5,17.7){\scalebox{1.1}{$\eta_\mathrm{d}$}}
\put(50.3,17.7){\scalebox{1.1}{$\eta_\mathrm{d}$}}
\put(85.3,17.7){\scalebox{1.1}{$\eta_\mathrm{d}$}}

\put(14.5,15.7){\scalebox{1.1}{$\eta_\mathrm{up}$}}
\put(50.3,15.7){\scalebox{1.1}{$\eta_\mathrm{up}$}}
\put(85.3,15.7){\scalebox{1.1}{$\eta_\mathrm{up}$}}

\put(14.3,13.3){{\scalebox{1}{$W_\mathrm{tot}^\mathrm{d}$}}}
\put(50,13.3){{\scalebox{1}{$W_\mathrm{tot}^\mathrm{d}$}}}
\put(85,13.3){{\scalebox{1}{$W_\mathrm{tot}^\mathrm{d}$}}}

\put(14.5,-1.5){\scalebox{1.1}{{$\Delta I$}}}
\put(48.7,-1.5){\scalebox{1.1}{{$\Delta I$}}}
\put(82.5,-1.5){\scalebox{1.1}{{$\Delta I$}}}
\end{overpic}
\caption{Performance characteristics of the spin-1/2 CIE and the standard Carnot engine. The absorbed heat from the cold reservoir $Q_\mathrm{c}^\mathrm{d}$, the total input energy $ Q_\mathrm{h}^\mathrm{d}$,  the total output work $ W_\mathrm{tot}^\mathrm{d}$,   the  information changes $\Delta I$ in the spin-1/2 CIE, the output work $W_\mathrm{tot}$  in the standard Carnot engine  as functions of the inverse hot temperature $\beta_\mathrm{h}$ for (a) $\epsilon=0.24$, (b) $\epsilon=0.28$, and (c) $\epsilon=0.32$. The output  work $W_\mathrm{tot}^\mathrm{d}$, the efficiency $\eta_\mathrm{d}$,  and the efficiency upper bound $\eta_\mathrm{up}$  as the functions of the information changes $\Delta I$ for (d) $\beta_\mathrm{h}=0.4$, (e) $\beta_\mathrm{h}=0.5$, and (f) $\beta_\mathrm{h}=0.6$.   The other parameters are $\omega_A=1.5$, $\omega_C=2$, $\beta_\mathrm{c}=1$, and $n_e=-n_g=1/2$.}
    \label{heatee}
\end{figure*}

\textit{The performance of the information Carnot engine}.$-$
The net work produced by the information-driven cycle is defined as 
\begin{eqnarray}\label{wnet}
 W_\mathrm{net}= W_\mathrm{d}-\beta_\mathrm{c}^{-1}\Delta I. 
\end{eqnarray}
Combining Eqs. (\ref{Qh}), (\ref{Wd}), (\ref{Qc1}), and (\ref{wnet}), the total output work produced by the working system's size variation after completing one  cycle is given by
\begin{eqnarray}\label{Wtot1} W_\mathrm{tot}^\mathrm{d}
=W_\mathrm{tot}+ W_\mathrm{net},
\end{eqnarray}  
where $ W_\mathrm{tot}= Q_\mathrm{h}+Q_\mathrm{c}=(S_\mathrm{h}-S_\mathrm{c})(\beta_\mathrm{h}^{-1}-\beta_\mathrm{c}^{-1})$ quantifies the output work of the standard Carnot cycle.  A key observation from Eq. (\ref{Wtot1}) is that when the net work $W_\mathrm{net}$ is positive, the CIE produces greater output work than the standard Carnot engine. Moreover, this reveals the first advantage of the CIE: in the regime $S_\mathrm{h}<S_\mathrm{c}<S_\epsilon$ ($S_\mathrm{c}<S_\epsilon$ ensures $Q_\mathrm{c}^\mathrm{d}<0$ ), the operation of the standard Carnot engine as a heat engine is forbidden due to  $ Q_\mathrm{c}>0$, $ Q_\mathrm{h}<0$, and $W_\mathrm{tot}<0$. However,  if the net work satisfies  $W_\mathrm{net}>(S_\mathrm{c}-S_\mathrm{h})(\beta_\mathrm{h}^{-1}-\beta_\mathrm{c}^{-1})$, where the net work generated by  information-driven cycle compensates for the negative work produced by the standard Carnot cycle,  the CIE can operate as a heat engine because the total work output  and the total energy injected into the system which is defined as $Q_\mathrm{h}^\mathrm{d}= Q_\mathrm{h}+W_\mathrm{d}$ of the CIE satisfy  $ W_\mathrm{tot}^\mathrm{d}>0$ and $Q_\mathrm{h}^\mathrm{d}>(S_\epsilon-S_\mathrm{c})\beta_\mathrm{c}^{-1}>0$.
 
The efficiency $\eta_\mathrm{d}$  (see Appendix \ref{defi})  of the CIE, combining Eqs. (\ref{Qh}), (\ref{Wd}), and (\ref{Wtot1}), is given by
\begin{equation}\label{etad}
\eta_\mathrm{d}=\frac{ W_\mathrm{tot}^\mathrm{d}}{Q_\mathrm{h}^\mathrm{d}}
    =\eta_\mathrm{C}+\frac{1}{\beta_\mathrm{c}}\frac{\beta_\mathrm{h} W_\mathrm{d}-\Delta I}{Q_\mathrm{h}^\mathrm{d}}.
\end{equation}
Eq. (\ref{etad}) indicates that the condition for the CIE efficiency to defeat the standard Carnot efficiency is that the work performed by the demon on the system surpasses the heat required to flow from the hot reservoir to the system to produce an equivalent entropy change as that generated by the work.
In Appendix \ref{bound}, we have proven that $\beta_\mathrm{h} W_\mathrm{d}\geq \Delta I\geq \beta_\mathrm{d}W_\mathrm{d}$. By incorporating this relationship into  Eq. (\ref{etad}), the efficiency bounds  can be obtained
\begin{eqnarray}\label{etadu}
\eta_\mathrm{C}\leq
\eta_\mathrm{d}\leq\eta_\mathrm{up}=\eta_\mathrm{C}+\frac{\beta_\mathrm{h}}{\beta_\mathrm{c}}\frac{ \Delta I\Delta T}{Q_\mathrm{h}^\mathrm{d}}, 
\end{eqnarray}
where $\Delta T=\beta_\mathrm{d}^{-1}-\beta_\mathrm{h}^{-1}$ denotes the temperature variation of the system, induced by the demon control during the stroke  $C\rightarrow C^{'}$.
The term $\eta_\mathrm{up}$ represents the upper bound of the CIE efficiency and serves, to some extent, as an indicator of the efficiency level. According to the definition of $ \Delta I$, the information changes increase with the measurement error within the regime  $0\leq \epsilon\leq 0.5$.
As the measurement error increases, the effective temperature
$\beta_\mathrm{d}^{-1}$ also rises, leading to a positive correlation between $\beta_\mathrm{d}^{-1}$ and $\Delta I$.  When $\Delta I=0$,  where $S_\epsilon=S_\mathrm{h}$, resulting in $\beta_\mathrm{d}^{-1}=\beta_\mathrm{h}^{-1}$.
Consequently,  if $\Delta I<0$, then  $\Delta T<0$; conversely,  if $\Delta I>0$, then  $\Delta T>0$. 
Thus, in the regime $\Delta I>0$, $\eta_\mathrm{up}$ increases with  $\Delta I$, while in the regime $\Delta I<0$, $\eta_\mathrm{up}$ increases as $\Delta I$ decreases.
Furthermore, when  $\Delta I=0$, the equation in Eq. (\ref{etadu}) will be 
satisfied: $\eta_\mathrm{d}=\eta_\mathrm{up}=\eta_\mathrm{C}$, reflecting the enhancement of CIE efficiency through the information changes.

The first inequality in Eq. (\ref{etadu}) reveals the second advantage of the CIE:  Its efficiency  is greater than or equal to  than the standard Carnot due to the information changes.
In Eqs. (\ref{Qc1}), the condition for the system to absorb zero heat from the cold reservoir is given by 
 $S_\epsilon=S_\mathrm{c}$, causing the third advantage of the CIE emerge.  Under this condition, the information changes become to $\Delta I=S_\mathrm{c}-S_\mathrm{h}$.  At this time, the total output work satisfies $W_\mathrm{tot}^\mathrm{d}=Q_\mathrm{h}^\mathrm{d}=W_\mathrm{d}-\Delta I/\beta_\mathrm{h}\geq0$, meaning the CIE efficiency  reaches 100\%, 
 which is forbidden for the standard Carnot engine.

\textit{Numerical analysis for spin-1/2  Carnot information engine}.$-$
To illustrate our ideas, we present a numerical analysis using a spin-1/2 CIE as depicted in Fig. \ref{model}(b),  where  the populations are distributed as  $n_e=-n_g=1/2$.
Figs. \ref{heatee}(a)-(c) show that the CIE exhibits an extended operational regime compared to the standard Carnot engine. The numerical results demonstrate that the CIE can operate as a heat engine  in the regime $\beta_\mathrm{c}\omega_A/\omega_C=0.75\leq\beta_\mathrm{h}<\beta_\mathrm{c}$, for $ Q_\mathrm{c}^\mathrm{d}<0, Q_\mathrm{h}^\mathrm{d}>0,$ and $ W_\mathrm{tot}^\mathrm{d}>0$.  However, in this regime, the standard Carnot heat engine is thermodynamically forbidden, for $W_\mathrm{tot}<0$. This extension stems from two mechanisms: (a) the net work generated by information changes, $ W_\mathrm{net}$,  offsets the negative work produced by the standard  Carnot cycle, thereby enabling the  CIE to perform positive work overall;  (b) The dissipated heat $\Delta I/\beta_\mathrm{c}$ caused by the information changes  counterbalance the positive heat  $Q_\mathrm{c}$, effectively resulting in a positive heat  $- Q_\mathrm{c}^\mathrm{d}$ flow into the cold reservoir.  Moreover, Figs. \ref{heatee}(a)-(c) reveal that in the regime of positive information changes ($\Delta I>0$), where the net work $W_\mathrm{net}>0$, the CIE  produces more work compared to the standard Carnot heat engine,   consistent with  Eq. (\ref{Wtot1}). 
The efficiency $\eta_\mathrm{d}$ as a function of $\beta_\mathrm{h}$ please see Appendix \ref{INC}.

In Figs. \ref{heatee}(d)-(f), the red dashed curve, while consistently remaining above the Carnot efficiency (horizontal axis), lies below the red dotted curve, confirming  Eq. (\ref{etadu}). Additionally, $\eta_\mathrm{d}$  exhibits the same behavior as $\eta_\mathrm{up}$ with increasing $\Delta I$. In the regime $\Delta I\leq0$,  $\eta_\mathrm{d}$ monotonically increases from the Carnot value $\eta_\mathrm{C}$
 as $\Delta I$ decreases, and reaches the perfect efficiency $\eta_\mathrm{d}=1$
 when $\Delta I$ reaches the minimum $\Delta I=S_\mathrm{c}-S_\mathrm{h}$. At this point, the CIE  absorbs zero heat from the cold reservoir, enabling complete conversion of the input energy $Q_\mathrm{h}^\mathrm{d}$ into output work. In the regime 
 $\Delta I\geq0$,
 the efficiency $\eta_\mathrm{d}$ increases monotonically with information changes until $\Delta I$ reaches its maximum at $ \Delta I=\ln2-S_\mathrm{c}$. 
 The analysis of this non-monotonic efficiency  behavior with respect to the information changes is provided in Appendix \ref{asd}.

\textit{Experimental scheme}.$-$The CIE  can be implemented experimentally  by  a
single trapped $^{40}\mathrm{Ca}^+$ ion, where  
 the $S_{1/2}$ and $S_{-1/2}$ energy levels  encode the system's states, while the ion's motional modes serve as the work  storage. The total Hamiltonian of the ion \cite{Lutz24,OV23} consists of three components  
$H=H_\mathrm{s}+H_\mathrm{har}+H_\mathrm{int}$, where  $H_\mathrm{s}=\omega_\mathrm{s}(t)\sigma_z/2+\Omega_\mathrm{s}\sigma_x/2$ represents the system spin Hamiltonian, $H_\mathrm{har}=\omega_\mathrm{har}a^{\dag}a$ describes  the  harmonic oscillator  Hamiltonian used for work storage, and  $H_\mathrm{int}=-(\tilde{\eta}\Omega_\mathrm{s}/2)\sin(\omega_\mathrm{har}t)\sigma_y$ serves as the interaction term. Here, $\omega_\mathrm{s}(t)$, $\Omega_\mathrm{s}$, $\omega_\mathrm{har}$,   $\tilde{\eta}$, $\sigma_{x,y,z}$, and $a^{\dag}$ ($a$) denote the  driving external field,  the Rabi frequency of the driving external field, the oscillator frequency, the Lamb-Dicke parameter, the Pauli matrices, and the  creation (annihilation) operators, respectively.
The adiabatic strokes are achieved through slow modulation of  $\omega_\mathrm{s}$ \cite{RJ19}.  The isothermal strokes are implemented by discretizing  it into  $N$  segments, combining adiabatic evolution and isochoric thermalization  \cite{Quan08,Kos92}.
Measurement and feedback control can be realized via two composite light fields, with system control based on fluorescence frequency detection. For more details, please see Appendix \ref{exper}.

\textit{Conclusions}.$-$Based on the standard Carnot cycle, we have developed an CIE, driven by  two thermal reservoirs and an information reservoir.    
We demonstrate that information changes enable the CIE to operate in thermodynamically forbidden regimes of the standard Carnot cycle, lead that the efficiency of the CIE is greater  than or equal to the standard Carnot efficiency. Most significantly, the CIE can achieve 100\% efficiency ($\eta_\mathrm{d}=1$) with positive work output. To further confirm our theoretical findings, we propose a spin-1/2 CIE as a practical example, and design an  experimental scheme for implementing the spin-1/2 CIE based on a trapped $^{40}\mathrm{Ca}^+$ ion system. 
Our findings not only provide a theoretical framework for designing information-assisted heat engines that transcend conventional thermodynamic bounds, but also pave the way for the practical realization of dissipationless heat engines.

\begin{acknowledgments}
 Y. Xiao thanks the support in part by the National Natural Science Foundation of China Grant No. 21721003. \end{acknowledgments}
\appendix
\begin{widetext}
\section{The equations of the measurement and control strokes}\label{inf}
At point $C$, following the measurement operation, the system-demon correlation is established. The mutual information between them is quantified as \cite{TM60}
\begin{eqnarray}\label{SIC}
I_C&=&\mathop{\sum}\limits_{x_C,y}P(y|x_C)P(x_C)\ln\frac{P(y|x_C)}{P(y)}\nonumber\\
    &=&P(y=e|x_C=e)P(x_C=e)\ln\frac{P(y=e|x_C=e)}{P_y^e}+P(y=g|x_C=e)P(x_C=e)\ln\frac{P(y=g|x_C=e)}{P^g_y}\nonumber\\
    &+&P(y=e|x_C=g)P(x_C=g)\ln\frac{P(y=e|x_C=g)}{P_y^e}+P(y=g|x_C=g)p(x_C=g)\ln\frac{P(y=g|x_C=g)}{P_y^g}\nonumber\\
    &=&(1-\epsilon)\ln(1-\epsilon)+\epsilon\ln \epsilon-[P_C^e(1-\epsilon)+P_C^g\epsilon]\ln P_y^e
    -[P_C^g(1-\epsilon)+P_C^e\epsilon]\ln P_y^g\nonumber\\
    &=&S_Y-S_\epsilon,
\end{eqnarray}
where, $\epsilon=P(y\neq x_C|x_C)$ denotes the measurement error.  $P_C^e=P(x_C=e)$, and $P_C^g=P(x_C=g)$ represent the probabilities of the system being in excited state and ground state at point $C$, respectively. Here   $S_Y=-P_y^g\ln P_y^g-P_y^e\ln P_y^e$ represents the demon's entropy, with $P^e_y=P(y=e)=P_C^e(1-\epsilon)+P_C^g\epsilon$ and $P^g_y=P(y=g)=P_C^e\epsilon+P_C^g(1-\epsilon)$ being the probabilities of measurement outcomes $e$ and $g$, respectively. $S_\epsilon=-(1-\epsilon)\ln(1-\epsilon)-\epsilon\ln\epsilon$ denotes the entropy of the measurement error. Then, the demon utilizes this information to perform work to modulate the system's state, which makes the system state become to  $P^e_{C^{'}}=
   P^e_CP(y=g|x_C=e)
   +P^g_CP(y=e|x_C=g)
   =\epsilon $ and $ P^g_{C^{'}}=P^e_CP(y=e|x_C=e)
  +P^g_CP(y=g|x_C=g)=1-\epsilon$. 
Based  on the distribution $P_{C^{'}}^e$ and  $P_{C^{'}}^g$,  the work exerted by the demon on the system  can be calculated as 
\begin{eqnarray}
 W_\mathrm{d}
&=&\mathop{\sum}\limits_{x_C,x_{C^{'}},y} P(x_C\rightarrow x_{C^{'}}|y)P(y)(\varepsilon_C^{x_{C^{'}}}-\varepsilon_C^{x_C})=\mathop{\sum}\limits_{x_C,x_{C^{'}}}P(x_C,x_{C^{'}})(\varepsilon_C^{x_{C^{'}}}-\varepsilon_C^{x_{C}})\nonumber\\
&=&\mathop{\sum}\limits_{x_{C^{'}}}\mathop{\sum}\limits_{x_{C}}P(x_C,x_{C^{'}})\varepsilon_C^{x_{C^{'}}}-\mathop{\sum}\limits_{x_{C}}\mathop{\sum}\limits_{x_{C^{'}}}P(x_C,x_{C^{'}})\varepsilon_C^{x_C}=\mathop{\sum}\limits_{x_{C^{'}}}P(x_{C^{'}})\varepsilon_C^{x_{C^{'}}}-\mathop{\sum}\limits_{x_{C}}P(x_C)\varepsilon_C^{x_C}\nonumber\\
&=&\omega_C[(P_{C^{'}}^en_e+P_{C^{'}}^gn_g)-(P_C^en_e+P_C^gn_g)],
\end{eqnarray}
where $P(x_C\rightarrow x_{C^{'}}|y)$ is the condition probability for the 
system's evolutionary trajectory from $x_C$ to $x_{C^{'}}$ based on   the measurement outcome $y$  which is equal to the condition probability $P(x_C|y)$ due to this control stroke is nondegenerate.
After the first control sroke $C\rightarrow C^{'}$, the residual mutual information takes the form:
\begin{eqnarray}\label{SIC1}
I_{C^{'}}&=&\mathop{\sum}\limits_{x_{C^{'}},y}P(x_{C^{'}}|y)P(y)\ln\frac{P(x_{C^{'}}|y)}{P(x_{C^{'}})}\nonumber\\
&=&P(x_{C^{'}}=e|y=e)P_y^e\ln\frac{P(x_{C^{'}}=e|y=e)}{P_{C^{'}}^e}+P(x_{C^{'}}=g|y=e)P_y^e\ln\frac{P(x_{C^{'}}=g|y=e)}{P_{C^{'}}^g}\nonumber\\
&+&P(x_{C^{'}}=e|y=g)P_y^g\ln\frac{P(x_{C^{'}}=e|y=g)}{P_{C^{'}}^e}+P(x_{C^{'}}=g|y=g)P_y^g\ln\frac{P(x_{C^{'}}=g|y=g)}{P_{C^{'}}^g}\nonumber\\
&=&P(y=e|x_C=g)P_C^g\ln\frac{P(y=e|x_C=g)P_C^g}{P_y^eP_{C^{'}}^e}
+P(y=e|x_C=e)P_C^e\ln\frac{P(y=e|x_C=e)P_C^e}{P_y^eP_{C^{'}}^g}\nonumber\\
&+&P(y=g|x_C=e)P_C^e\ln\frac{P(y=g|x_C=e)P_C^e}{P_y^gP_{C^{'}}^e}
+P(y=g|x_C=g)P_C^g\ln\frac{P(y=g|x_C=g)P_C^g}{P_y^gP_{C^{'}}^g}\nonumber\\
&=&P_C^e\ln P_C^e+P_C^g\ln P_C^g-[P_C^e(1-\epsilon)+P_C^g\epsilon]\ln P_y^e-[P_C^g(1-\epsilon)+P_C^e\epsilon]\ln P_y^g.
\end{eqnarray}
 Following \cite{TC00,Zeng21},  through the stochastic entropy flows from the system, the heat absorbed from the cold reservoir during the third control stroke can be calculated as 
\begin{eqnarray}\label{SQXY}
 Q_\mathrm{c}^\mathrm{d} &=&\mathop{\sum}\limits_{x_{D^{'}}x_A,y}\beta_\mathrm{c}^{-1}P({x}_{D^{'}}\rightarrow x_A|x_{D^{'}},y)P(x_{D^{'}}|y)P(y)\ln\frac{P({x}_A\rightarrow x_{D^{'}}|x_A,y)}{P({x}_{D^{'}}\rightarrow x_A|x_{D^{'}},y)}\nonumber\\
&=&\beta_\mathrm{c}^{-1}\mathop{\sum}\limits_{x_{D^{'}},x_A,y}P(x_A)P(x_{D^{'}}|y)P(y)\ln\frac{P(x_{D^{'}})}{P(x_A)}\nonumber\\
&=&\beta_\mathrm{c}^{-1}\mathop{\sum}\limits_{x_A}P(x_{A})\mathop{\sum}\limits_{x_{D^{'}}}\mathop{\sum}\limits_{y}P(x_{D^{'}},y)\ln P(x_{D^{'}})-\beta_\mathrm{c}^{-1}\mathop{\sum}\limits_{x_{D^{'}},y}P(x_{D^{'}},y)\mathop{\sum}\limits_{x_A}P(x_{A})\ln P(x_A)\nonumber\\
&=&\beta_\mathrm{c}^{-1}[\mathop{\sum}\limits_{x_{D^{'}}}P(x_{D^{'}})\ln{P(x_{D^{'}})}-\mathop{\sum}\limits_{x_A}P(x_A)\ln{P(x_A)}]\nonumber\\
&=&\beta_\mathrm{c}^{-1}(S_\mathrm{c}-S_\epsilon),
\end{eqnarray}
which is inequivalent to  $\beta_\mathrm{c}^{-1}\langle \Delta S_{X|Y}^{D^{'}A}\rangle$. Here, $\langle \Delta S_{X|Y}^{D^{'}A}\rangle$ represents the average conditional entropy production along the trajectory 
  $x_{D^{'}}\rightarrow x_A|y$, given by 
\begin{eqnarray}\label{DSXY}
\langle \Delta S_{X|Y}^{D^{'}A}\rangle&=&\mathop{\sum}\limits_{x_D^{'},x_A,y}P(x_{D^{'}}\rightarrow A|y)P(y)\ln\frac{P(x_{D^{'}}|y)}{P(x_A|y)}=\mathop{\sum}\limits_{x_D^{'},x_A,y}P(x_{D^{'}}|y)P(y)P(x_A)\ln\frac{P(x_{D^{'}}|y)}{P(x_A)}\nonumber\\
&=&\mathop{\sum}\limits_{x_D^{'},x_A,y}P(x_{D^{'}},y)P(x_A)\ln\frac{P(x_{D^{'}}|y)}{P(x_A)}
=\mathop{\sum}\limits_{x_D^{'},x_A,y}P(x_{D^{'}},y)P(x_A)[\ln\frac{P(x_{D^{'}}|y)}{P(x_{D^{'}})}+\ln\frac{P(x_{D^{'}})}{P(x_A)}]\nonumber\\
&=&\mathop{\sum}\limits_{x_A}P(x_A)\mathop{\sum}\limits_{x_D^{'},y}P(x_{D^{'}},y)\ln\frac{P(x_{D^{'}}|y)}{P(x_{D^{'}})}+\mathop{\sum}\limits_{x_A}P(x_A)\mathop{\sum}\limits_{x_D^{'}}\ln P(x_{D^{'}})\mathop{\sum}\limits_{y}P(x_{D^{'}},y)\nonumber\\
&-&\mathop{\sum}\limits_{x_D^{'},y}P(x_{D^{'}},y)\mathop{\sum}\limits_{x_A}P(x_A)\ln P(x_A)\nonumber\\
&=&\mathop{\sum}\limits_{x_D^{'},y}P(x_{D^{'}},y)\ln\frac{P(x_{D^{'}}|y)}{P(x_{D^{'}})}+\mathop{\sum}\limits_{x_D^{'}}P(x_{D^{'}})\ln P(x_{D^{'}})
-\mathop{\sum}\limits_{x_A}P(x_A)\ln P(x_A)\nonumber\\
&=&S_\mathrm{c}-S_\epsilon+ I_{C^{'}},
\end{eqnarray}
where  we have used $P(x_{D^{'}}=e)=P(x_{C^{'}}=e)=\epsilon$ and $P(x_{D^{'}}=g)=P(x_{C^{'}}=g)=1-\epsilon$. Here, $S_\mathrm{c}=-\mathop{\sum}\limits_{x_A}P(x_A)\ln P(x_A)=-P_A^e\ln P_A^e-P_A^g\ln P_A^g$ denotes the system's entropy at point $A$, while $S_\epsilon=-\mathop{\sum}\limits_{x_{D^{'}}}P(x_{D^{'}})\ln P(x_{D^{'}})=-\epsilon\ln\epsilon-(1-\epsilon)\ln(1-\epsilon)$ represents the  entropy of the system at point $D^{'}$ which is also the  entropy of measurement error. In Eq. (\ref{SQXY}), the equivalence between the absorbed heat from the cold reservoir and the product of the cold temperature and system's coarse-grained entropy change indicates reversible dynamics at the coarse-grained level for the composite system (system+reservoir). The total microscopic entropy production, including the system, the cold reservoir, and the demon, is defined as  $ \sigma_{X|Y}=\Delta S_{X|Y}^{D^{'}A}-\beta_\mathrm{c} Q_c^\mathrm{d}$ which satisfies $\sigma_{X|Y}= I_{C^{'}}\neq0$, revealing irreversible dynamics in the complete system (system+reservoir+demon) due to information dissipation \cite{Zeng21}.

\section{The definition of the  CIE efficiency}\label{defi}
We present some fundamental justifications for defining the efficiency of the Carnot information engine (CIE) as the ratio between the work output of the working system and the total energy investment.   (i) Considering  the  definition where total output work comprises the demon-extracted work, $- W_\mathrm{d}$, and working system  producing,  $ W^\mathrm{d}_\mathrm{tot}$, which reads 
  $ W=W^\mathrm{d}_\mathrm{tot}-W_\mathrm{d}=Q_\mathrm{c}^\mathrm{d}+Q_\mathrm{h}$ \cite{Lutz23},  
the corresponding efficiency  $\eta= W /Q_\mathrm{h}=\eta_\mathrm{C}-\Delta I/(\beta_\mathrm{c}Q_h)$ coincides with the upper bound derived for discrete quantum measurement and control \cite{Sagawa08} when $ I_{C^{'}}=0$. Obviously, this efficiency can exceed the Carnot limit ($\eta>\eta_\mathrm{C}$)  when $\Delta I<0$. In the special case where $Q_\mathrm{c}^\mathrm{d}=0$, we obtain  $W=Q_\mathrm{h}\neq0$, achieving perfect efficiency $\eta=1$. Thus, even under this  definition,  in our CIE the information changes  still enable it to defeat the standard  Carnot limit. (ii) The working system and the demon operate through fundamentally distinct mechanisms in work generation. The working system produces work through size changes, analogous to volume changes in the classical steam engines. In contrast, the demon extracts work by modifying the system's distribution—a qualitatively different process that alters the statistical properties rather than mechanical coordinates of the system. Therefore, combining the demon-extracted work and working system output work into a total work output lacks physical justification, for their fundamentally distinct operational mechanisms.
 (iii) The work done  on the system by the demon $W_\mathrm{d}$ is used to  change the distribution of the system with constant frequency, which is similar to inject equivalent heat into the system. Additionally, in Refs. \cite{Nie16,Nie18}, the heat engine efficiency is defined as the ratio of the work produced by the working system to the total invested energy.  The total invested energy consists of the  passive thermal energy absorbed from the equilibrium reservoir and the ergotropy defined as the work required to transition the system from a passive to a non-passive state. Defining the total injection  energy of the CIE as the sum of the absorbed heat from the hot reservoir and the work exerted by the demon  is same as the total invested energy in Refs. \cite{Nie16,Nie18}. 
 Thus, defining the CIE's efficiency as $ W_\mathrm{tot}^\mathrm{d}/( Q_\mathrm{h}+W_\mathrm{d})$ is well-founded and physically justified. Furthermore, this definition, treating the demon's work as an energy investment rather than output, provides a more physically meaningful characterization of the engine's performance.

\section{The proof of the CIE efficiency bounds}\label{bound}
Considering an isochoric process in which the system's frequency is held constant at $\omega_C$.  The initial state distribution of the system mirrors the distribution observed at point $C^{'}$ ($C$). Subsequently, the system interacts weakly with a thermal reservoir characterized by an inverse temperature $\beta_\mathrm{h}$ ($\beta_\mathrm{d}$). Over a sufficiently long time, this interaction drives the system into an equilibrium state corresponding to the distribution at point $C$ ($C^{'}$). During this process, the heat absorbed by the system is $- W_\mathrm{d}$ ($W_\mathrm{d}$) and the associated entropy change is $-\Delta I=S_\mathrm{h}-S_\epsilon$  ($\Delta I=S_\epsilon-S_\mathrm{h}$). According to the second law of  thermodynamics, the total entropy production satisfies $S_\mathrm{h}-S_\epsilon+\beta_\mathrm{h}  W_\mathrm{d}\geq0$ ($S_\epsilon-S_\mathrm{h}-\beta_\mathrm{d} W_\mathrm{d}\geq0$), resulting the information changes and the work accompanying it satisfy $\beta_\mathrm{h} W_\mathrm{d}\geq  \Delta I$ ($ \Delta I\geq \beta_\mathrm{d} W_\mathrm{d}$).

\section{The CIE EFFICIENCY AS A FUNCTION OF THE Hot Inverse Temperature}\label{INC}
\begin{figure*}
\centering
\begin{overpic}[width=17cm]{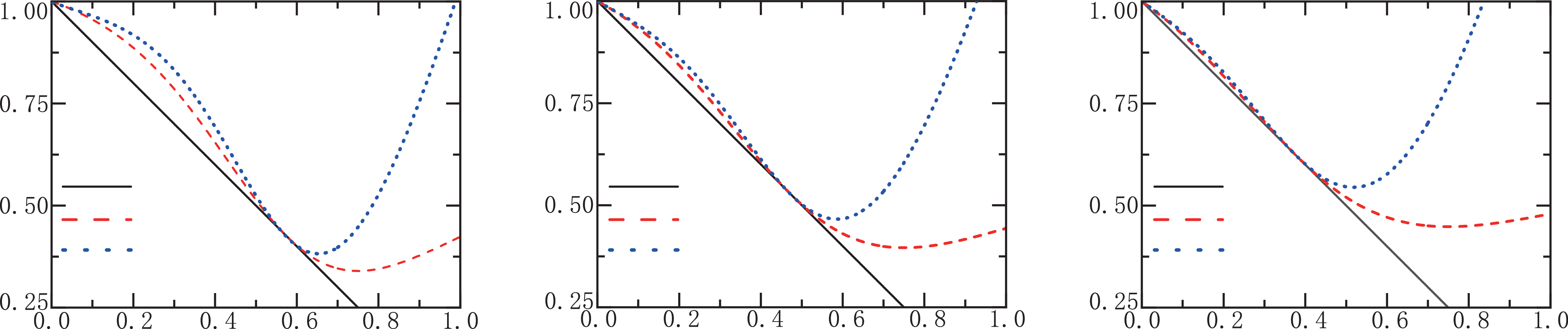}
\put(3,21.8){\scalebox{1.2}{(a)}}
\put(6.1,21.8){\scalebox{1.2}{$\epsilon=0.24$}}
\put(37.8,21.8){\scalebox{1.2}{(b)}}
\put(41,21.8){\scalebox{1.2}{$\epsilon=0.28$}}
\put(72.3,21.8){\scalebox{1.2}{(c)}}
\put(75.4,21.8){\scalebox{1.2}{$\epsilon=0.32$}}

\put(15.5,-0.9){\scalebox{1.2}{$\beta_\mathrm{h}$}}
\put(49.7,-0.9){\scalebox{1.2}{$\beta_\mathrm{h}$}}
\put(83.5,-0.9){\scalebox{1.2}{$\beta_\mathrm{h}$}}

\put(8.7,8.8){\scalebox{1.2}{$\eta_\mathrm{C}$}}
\put(43.5,8.8){\scalebox{1.2}{$\eta_\mathrm{C}$}}
\put(78.2,8.8){\scalebox{1.2}{$\eta_\mathrm{C}$}}

\put(8.7,6.8){\scalebox{1.2}{$\eta_\mathrm{d}$}}
\put(43.5,6.8){\scalebox{1.2}{$\eta_\mathrm{d}$}}
\put(78.2,6.8){\scalebox{1.2}{$\eta_\mathrm{d}$}}

\put(8.7,4.8){\scalebox{1.2}{$\eta_\mathrm{up}$}}
\put(43.5,4.8){\scalebox{1.2}{$\eta_\mathrm{up}$}}
\put(78.2,4.8){\scalebox{1.2}{$\eta_\mathrm{up}$}}

\end{overpic}
    \caption{The standard Carnot efficiency, the CIE efficiency $\eta_\mathrm{d}$, and the upper bound of the CIE efficiency $\eta_\mathrm{up}$ as functions of the hot  inverse temperature $\beta_\mathrm{h}$ for (a) $\epsilon=0.24$, (b) $\epsilon=0.28$, and (c) $\epsilon=0.32$. The other parameters are $\omega_A=1.5$, $\omega_C=2$, $\beta_\mathrm{c}=1$, and $n_e=-n_g=1/2$}
    \label{ill}
\end{figure*}
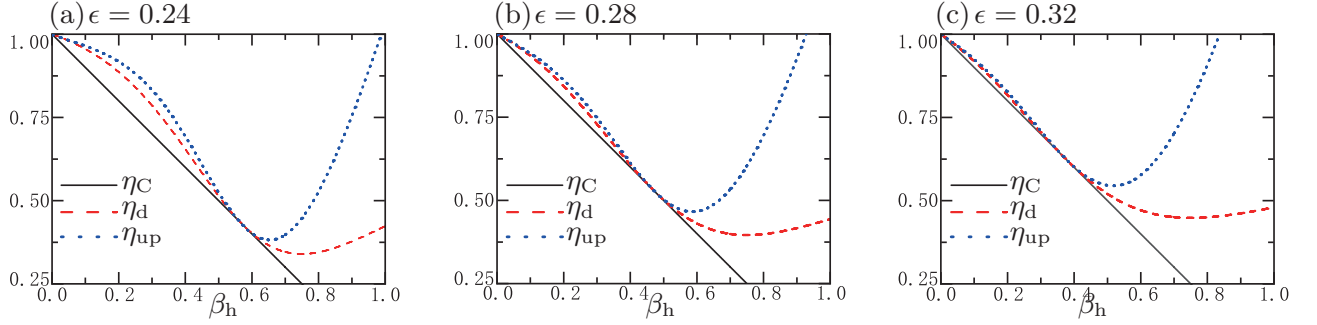
Figs. \ref{ill}(a)-(c) show that  the efficiency of the CIE increases and then decreases as $\beta_\mathrm{h}$ increases. The reason for this trend can be explained as:

Based on Eqs. (1), (3), and (5) in the main text, the total output work of the CIE is derived
\begin{eqnarray}\label{SWd}
 W_\mathrm{tot}^\mathrm{d}=(S_\mathrm{h}-S_\mathrm{c})(\frac{1}{\beta_\mathrm{h}}-\frac{1}{\beta_\mathrm{c}})+\omega_C (\langle n_\mathrm{d}\rangle- \langle n_\mathrm{h}\rangle)-\Delta I\beta_\mathrm{c}^{-1}.
\end{eqnarray}
Taking the  derivative of the output $W_\mathrm{tot}^\mathrm{d}$  with the hot inverse temperature $\beta_\mathrm{h}$, we derive
\begin{eqnarray}\label{dSWd}
(W_\mathrm{tot}^\mathrm{d})^{'}
 &=&   \frac{d W_\mathrm{tot}^\mathrm{d}}{d\beta_\mathrm{h}}
 =\frac{\frac{[1+(1+\beta_\mathrm{h}\omega_C)\exp(\beta_\mathrm{h}\omega_C)]\ln P^e_C+\exp(\beta_\mathrm{h}\omega_C)\{(\beta_\mathrm{h}\omega_C)^2+[1+\exp(\beta_\mathrm{h}\omega_C)-\beta_\mathrm{h}\omega_C]\ln P_C^g\}}{[1+\exp(\beta_\mathrm{h}\omega_C)]^2}+S_\mathrm{c}}{\beta_\mathrm{h}^2}.\nonumber\\
\end{eqnarray}
Given that  $\beta_\mathrm{h}^2>0$, the sign of the second  fraction in  Eq. (\ref{dSWd}) depends on the numerator $M_W$. To analyze the behavior of $M_W$ as a function of $\beta_\mathrm{h}$, we differentiate it with respect to  $\beta_\mathrm{h}$ and obtain 
\begin{eqnarray}\label{dMW}
   \frac{dM_W}{d\beta_\mathrm{h}}=\frac{1}{4}\beta_\mathrm{h}\omega_C^2\mathrm{sech}^2(\frac{\beta_\mathrm{h}\omega_C}{2})>0,
\end{eqnarray}
which reveals that  $M_W$ is an increasing function of $\beta_\mathrm{h}$.
 Noting that, when $\beta_\mathrm{h}=\beta_\mathrm{c}\omega_A/\omega_C$, $M_W=0$. It follows that  $M_W\leq0$ for $0<\beta_\mathrm{h}\leq\beta_\mathrm{c}\omega_A/\omega_C$, and $M_W>0$ for $\beta_\mathrm{c}>\beta_\mathrm{h}>\beta_\mathrm{c}\omega_A/\omega_C$, indicating that   $(W_\mathrm{tot}^\mathrm{d})^{'}\leq0$ in the regime $0<\beta_\mathrm{h}\leq\beta_\mathrm{c}\omega_A/\omega_C$, and $(W_\mathrm{tot}^\mathrm{d})^{'}>0$ in the regime $\beta_\mathrm{c}>\beta_\mathrm{h}>\beta_\mathrm{c}\omega_A/\omega_C$. Consequently,
the work  $W_\mathrm{tot}^\mathrm{d}$  in the regime $0<\beta_\mathrm{h}<\beta_\mathrm{c}\omega_A/\omega_C$  decreases 
and in the regime $\beta_\mathrm{c}>\beta_\mathrm{h}>\beta_\mathrm{c}\omega_A/\omega_C$   increases as $\beta_\mathrm{h}$ increases. According to Eq. (\ref{SQXY}), the heat absorbed from the cold reservoir $ Q_\mathrm{c}^\mathrm{d}=(S_c-S_\epsilon)/\beta_\mathrm{c}$ remains independent of 
 $\beta_\mathrm{h}$. Using the function $W_\mathrm{tot}^\mathrm{d}=Q_\mathrm{c}^\mathrm{d}+Q_\mathrm{h}^\mathrm{d}$, the efficiency  Eq. (8) in main text can be written as $\eta_\mathrm{d}= W_\mathrm{tot}^\mathrm{d}/( W_\mathrm{tot}^\mathrm{d}- Q_{c}^\mathrm{d})$, indicating
$\eta_\mathrm{d}$ is the monotonously increasing function of $W_\mathrm{tot}^\mathrm{d}$ as long as $Q_\mathrm{c}^\mathrm{d}$ is a constant. Thus,  in  Figs. \ref{ill}(a)-(c),  the  efficiency $\eta_\mathrm{d}$  of the CIE has  the same behavior as the work $W_\mathrm{tot}^\mathrm{d}$, initially decreasing and then increasing with $\beta_\mathrm{h}$. At $\beta_\mathrm{h}=\beta_\mathrm{c}\omega_A/\omega_C$,  where $ Q_\mathrm{h}=0$, the efficiency $\eta_\mathrm{d}$ reaches its minimum value $\eta_\mathrm{d}= W_\mathrm{net}/W_\mathrm{d}$, corresponding to the efficiency of the cycle $ C\rightarrow C^{'}\rightarrow D^{'}\rightarrow D\rightarrow C$ which is solely caused by the information changes.

Moreover, Figs. (\ref{ill})(a)-(c)  demonstrate that when information changes vanish, where the demon fails to exploit information for entropy manipulation, the CIE's efficiency coincides with the standard Carnot efficiency. However,  for non-zero information changes ($\Delta I\neq0$), regardless of whether the demon increases or reduces the system entropy, the CIE achieves higher efficiency than the Carnot limit,  as predicted by Eq. (9) in the main text.

\section{The CIE EFFICIENCY AS A FUNCTION OF the Information changes}\label{asd}
We now analyze how $\eta_\mathrm{d}$ varies with the information changes  $ \Delta I$.
From the relation $\langle \Delta n_\mathrm{d}\rangle=(2\epsilon-1)/2$, the average population $\langle n_\mathrm{d}\rangle$ satisfy the relation $-0.5\leq\langle n_\mathrm{d}\rangle\leq0$.
We express the measurement error in terms of $\langle n_\mathrm{d}\rangle$ as $\epsilon=(1+2\langle n_\mathrm{d}\rangle)/2$. Consequently, the measurement error entropy $S_\epsilon$ takes the form   $S_\epsilon=-(1+2\langle n_\mathrm{d}\rangle)/2\ln[(1+2\langle n_\mathrm{d}\rangle)/2]-(1-2\langle n_\mathrm{d}\rangle)/2\ln[(1-2\langle n_\mathrm{d}\rangle)/2]$,  which increases as  $\langle n_\mathrm{d}\rangle$ increases in the interval $-0.5\leq\langle n_\mathrm{d}\rangle\leq0$. Since $S_\mathrm{h}$ remains independent of the $\langle n_\mathrm{d}\rangle$, the information changes $\Delta I=S_\epsilon-S_\mathrm{h}$ also increase monotonically as  $\langle n_\mathrm{d}\rangle$ increases. 

According to Eq. (8) in main text, the derivative of $\eta_\mathrm{d}$ with respect to $\langle n_\mathrm{d}\rangle$ reads
\begin{eqnarray}\label{deta}
    \eta_\mathrm{d}^{'}=\frac{d\eta_\mathrm{d}}{d\langle n_\mathrm{d}\rangle}=\frac{\beta_\mathrm{h}\{\beta_\mathrm{h}\omega_C[2\ln2-2S_\mathrm{c}-\ln(1-2\langle n_\mathrm{d}\rangle)-\ln(1+2\langle n_\mathrm{d}\rangle)]-4\mathrm{arctanh}(2\langle n_\mathrm{d}\rangle)(\beta_\mathrm{h}\omega_C\langle n_\mathrm{h}\rangle+S_\mathrm{c}-S_\mathrm{h})\}}{2\beta_\mathrm{c}[\beta_\mathrm{h}\omega_C(\langle n_\mathrm{d}\rangle-\langle n_\mathrm{h}\rangle)+S_\mathrm{h}-S_\mathrm{c}]^2}.\nonumber\\
\end{eqnarray}
For notational convenience, we denote the numerator in the right-hand side of Eq. (\ref{deta}) as $M_e$. Given the positive denominator, the sign of Eq. (\ref{deta}) is determined solely by  $M_e$. To determine the sign of the numerator $M_e$, we calculate its derivative with respect to $\langle n_\mathrm{d}\rangle$ and obtain 
\begin{eqnarray}\label{ddeta}
    \frac{dM_e}{d\langle n_\mathrm{d}\rangle}=\frac{8\beta_\mathrm{h}^2[\omega_C(\langle n_\mathrm{d}\rangle)-\langle n_\mathrm{h}\rangle)+(S_\mathrm{h}-S_\mathrm{c})/\beta_\mathrm{h}]}{1-4\langle n_\mathrm{d}\rangle^2}.\nonumber\\
\end{eqnarray}
Given the constraints $-0.5\leq\langle n_\mathrm{d}\rangle\leq0$ and $\langle Q_\mathrm{h}^\mathrm{d}\rangle=\beta_\mathrm{h}\omega_C(\langle n_\mathrm{d}\rangle)-\langle n_\mathrm{h}\rangle)+(S_\mathrm{h}-S_\mathrm{c})/\beta_\mathrm{h}>0$, we find   $dM_e/\langle n_\mathrm{d}\rangle>0$, demonstrating that $M_e$ monotonically increases with $\langle n_\mathrm{d}\rangle$ increases. Because when $\langle n_\mathrm{d}\rangle=\langle n_\mathrm{h}\rangle$, $M_e=0$,  in the regime $\langle n_\mathrm{d}\rangle \leq\langle n_\mathrm{d}\rangle$, $M_e\leq 0$, and in the regime $\langle n_\mathrm{d}\rangle >\langle n_\mathrm{h}\rangle$, $M_e>0$. These relations indicate that $\eta_\mathrm{d}^{'}\leq 0$ in the regime $\langle n_\mathrm{d}\rangle \leq\langle n_\mathrm{d}\rangle$,  and $\eta_\mathrm{d}^{'}>0$ in the regime $\langle n_\mathrm{d}\rangle>\langle n_\mathrm{d}\rangle$.
So,  in the regime $\langle n_\mathrm{d}\rangle\leq\langle n_\mathrm{h}\rangle$, the efficiency $\eta_\mathrm{d}$ increases as $\langle n_\mathrm{d}\rangle$ decreases, and  in the regime $\langle n_\mathrm{d}\rangle>\langle n_\mathrm{h}\rangle$,  the efficiency $\eta_\mathrm{d}$ increases as $\langle n_\mathrm{d}\rangle$ increases. 
Because the information changes $\langle \Delta I\rangle$ also increase monotonically with $\langle n_\mathrm{d}\rangle$, and $\Delta I=0$ when $\langle n_\mathrm{d}\rangle=\langle n_\mathrm{h}\rangle$, the information changes satisfy the relations: If $\langle n_\mathrm{d}\rangle\leq\langle n_\mathrm{h}\rangle$, $\Delta I\leq 0$; If $\langle n_\mathrm{d}\rangle>\langle n_\mathrm{h}\rangle$, $\Delta I>0$.
Thus, the CIE efficiency decreases  in the  regime $\Delta I<0$, but increase in the  regime $\Delta I\geq0$ as the information changes increase.

\section{The EXPERIMENTAL SCHEME For The CIE}\label{exper}
We implement our  CIE using the electronic structure of a single trapped $^{40}\mathrm{Ca}^{+}$. Specifically, 
we use the $S_{1/2}$ and $S_{-1/2}$ energy levels  as the spin-up and spin-down states of the system. This spin system is coupled to a battery  for work storage, realized through the quantized harmonic motion of the trapped ion. The total Hamiltonian of the composite system comprises three terms \cite{Lutz24,OV23}: 
$H=H_\mathrm{s}+H_\mathrm{har}+H_\mathrm{int}$, where $H_\mathrm{s}=\omega_\mathrm{s}(t)\sigma_z/2+\Omega_\mathrm{s}\sigma_x/2$ is the Hamiltonian of the system with Pauli matrices $\sigma_{x,y,z}$. Here, $\omega_\mathrm{s}(t)$ denotes a time-dependent driving field that enables precise manipulation of the system  energy and $ \Omega_\mathrm{s}$ represents the Rabi frequency of the driving field.  
$H_\mathrm{har}=\omega_\mathrm{har}a^{\dag}a$ represents  the Hamiltonian of the battery where $\omega_\mathrm{har}$ 
  is the oscillator frequency, and $a^\dag$  
($a$) is the  creation (annihilation) operator of the motional modes.   $H_\mathrm{int}=-(\tilde{\eta}\Omega_\mathrm{s}/2)\sin(\omega_\mathrm{har}t)\sigma_y$ describes the system-battery interaction  where   $\tilde{\eta}$ represents the Lamb-Dicke parameter.

The adiabatic compression  stroke 
$A\rightarrow B$  is implemented by applying a time-dependent external driving field over a duration   $\tau_\mathrm{ch}$. The field varies linearly according to the function  $\omega_\mathrm{s}^\mathrm{ch}(t)=\omega_A(1-t/\tau_\mathrm{ch})+\omega_Bt/\tau_\mathrm{ch}$,  where $0\leq t\leq \tau_\mathrm{ch}$. During this process,  the system Hamiltonian does not commute with itself at different times, leading to nonadiabatic transitions. The probability of the system transitioning from state $|n\rangle$ to state $|m\rangle$ is given by $\xi=|\langle n|U_\mathrm{ch}|m\rangle|^2$ \cite{{JP29}}  with unitary operator $U_\mathrm{ch}=\mathcal{T}_{>}\exp[-i\int_0^{\tau}dtH_\mathrm{ch}(t)]$,  where $\mathcal{T}_{>}$ is the time-ordering operator and $H_\mathrm{s}^\mathrm{ch}=\omega_\mathrm{s}^\mathrm{ch}\sigma_z/2+\Omega_\mathrm{s}\sigma_x/2$. If the system is driven slowly, the condition 
  $\xi\rightarrow 0$ will be satisfied, representing the quantum adiabatic theorem.

\begin{figure*}
\centering
\begin{overpic}[width=8cm]{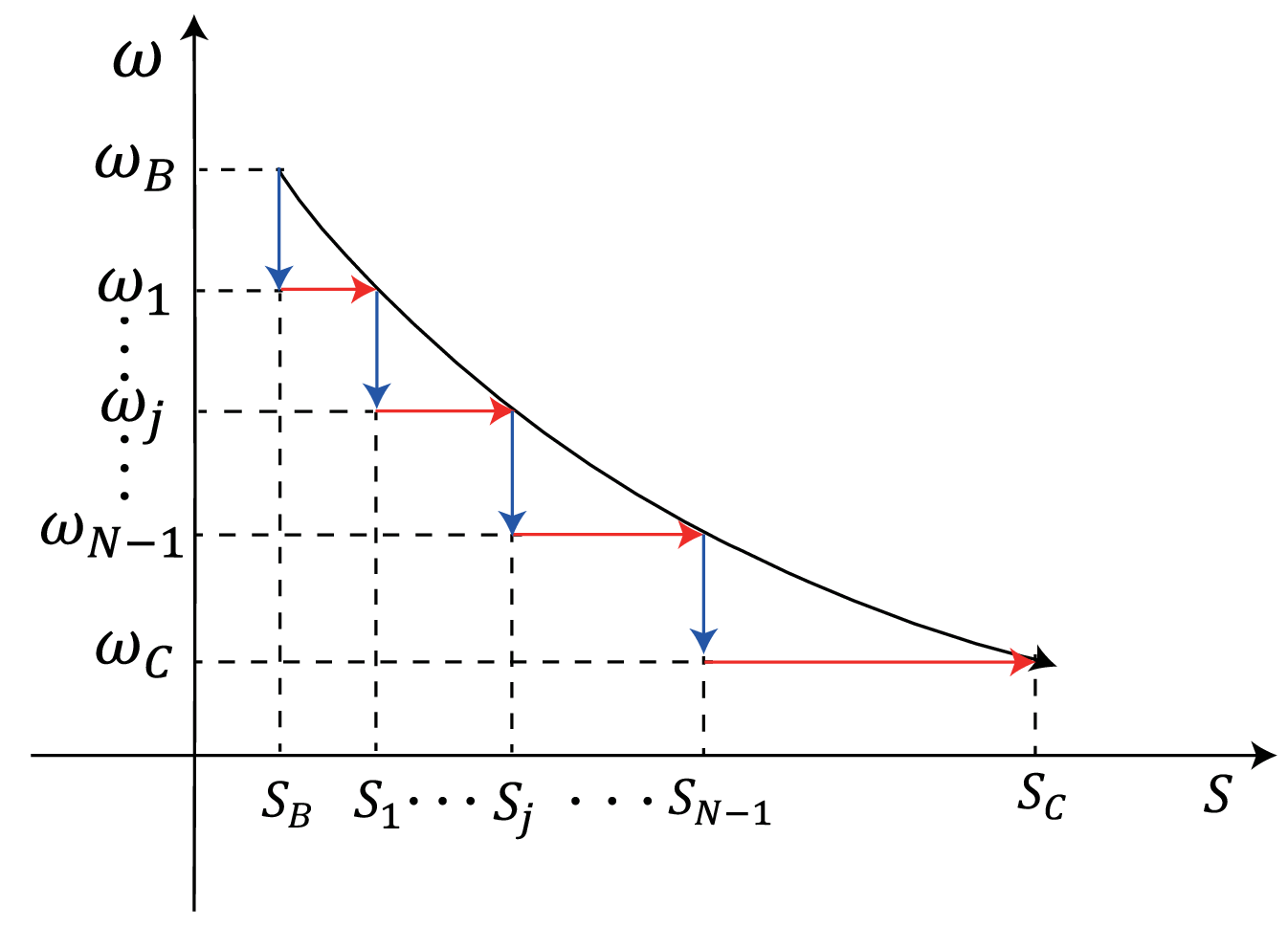}
\put(00,65){\scalebox{1.2}{(a)}}
\end{overpic}
\begin{overpic}[width=8cm]{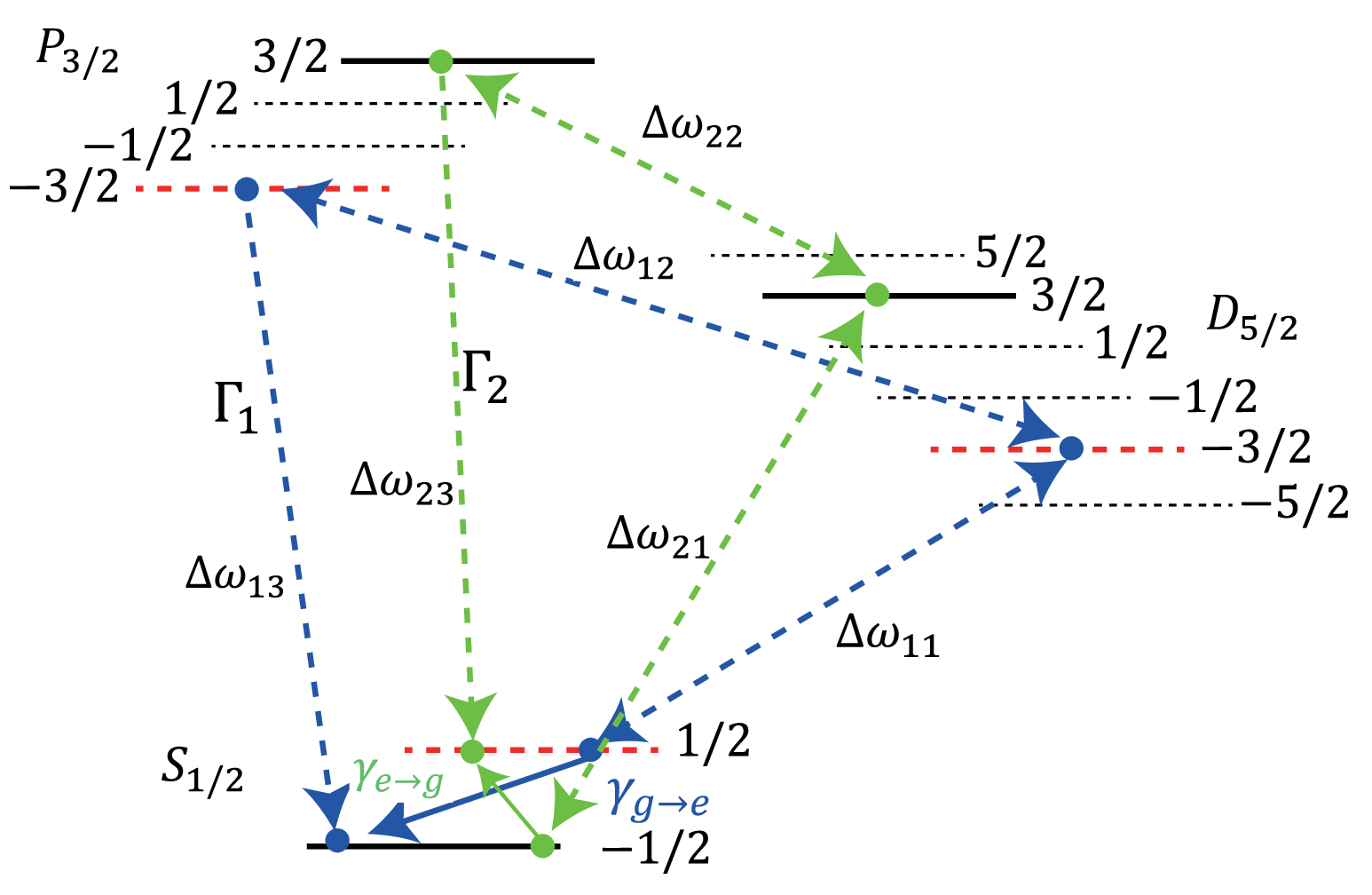}
\put(00,65){\scalebox{1.2}{(b)}}
\end{overpic}
    \caption{(a) Schematic illustration for discretization of isothermal expansion process $B\rightarrow C$.  Here the horizontal axis $S$ denotes the system entropy and the vertical axis $\omega$ represents the  level spacing of the spin-1/2 system. (b) Schematic illustration of $^{40}\mathrm{Ca}^{+}$ energy level and energy transitions.}
    \label{mmodel}
\end{figure*}
During the isothermal expansion stroke  $B\rightarrow C$,    the external field varies from $\omega_B$ to $\omega_C$, while maintaining thermal equilibrium at inverse temperature $\beta_\mathrm{h}$. To achieve this, we discretize this process into $N$ sequential segments, each comprising an adiabatic evolution and  an isochoric process \cite{Quan08,Kos92}, as shown in Fig. \ref{mmodel}(a). The field increment per segment is  $d\omega=(\omega_C-\omega_B)/N$.
For the $j$-th segment ($j=1,2,3,...N$),  the field evolves adiabatically  as  $\omega_\mathrm{s}^j(t)=\omega_{j-1}[1-(t-t_{j-1})/(t_{j}-t_{j-1})]+\omega_{j}(t_{j}-t)/(t_{j}-t_{j-1})$,
where $t_{j-1}\leq t\leq t_j$,  and $\omega^j=\omega_B+jd\omega$ represents the field at the $j$-th isochoric process.  In the isochoric process,  the thermalization is achieved through a laser-induced cascade process, involving multiple electronic levels of the calcium ion.
Specifically, to induce transitions from the excited (ground) state to the ground (excited) state, 
a laser with frequency $\Delta\omega_{11}^{j}$ ($\Delta\omega_{21}^{j}$ ) excites the electron from  $S_{1/2}$  ($S_{-1/2}$) to $D_{-3/2}$ ($D_{3/2}$) followed by another laser with frequency 
$\Delta\omega_{12}^{j}$  ($\Delta\omega_{22}^{j}$), driving the transition from $D_{-3/2}$ ($D_{3/2}$) 
  to  $P_{-3/2}$ ($P_{3/2}$). Here, $\Delta\omega_{11}^j, \Delta\omega_{12}^j, \Delta\omega_{21}^j$ and $\Delta\omega_{22}^j$ correspond to the energy differences between $S_{1/2}$ and $D_{-3/2}$, $D_{-3/2}$ and $P_{-3/2}$, $S_{-1/2}$ and $D_{3/2}$, and $D_{3/2}$ and $P_{3/2}$.  The electron then decays radiatively to  $S_{-1/2}$ ($S_{1/2}$) with decay rate $\Gamma_1$ ($\Gamma_2$), as shown in Fig. \ref{mmodel}(b). The effective  transition rates from $S_{1/2}$  to $S_{-1/2}$  and  from $S_{-1/2}$ to $S_{1/2}$  after these transitions are given by \cite{Yan22}
  \begin{eqnarray}
      \gamma_{e\rightarrow g}^j=(\Omega_{11}^j)^2\Gamma_1/(\Omega_{12}^{j})^2, \gamma_{g\rightarrow e}^j=(\Omega_{21}^j)^2\Gamma_2/(\Omega_{22}^{j})^2,
  \end{eqnarray}
 where $\Omega_{11}^j, \Omega_{12}^j, \Omega_{21}^j$,  and $\Omega_{22}^j$ are the Rabi frequencies corresponding to the laser frequencies $\Delta\omega_{11}^j, \Delta\omega_{12}^j, \Delta\omega_{21}^j$ and $\Delta\omega_{22}^j$,  respectively. To ensure thermal equilibrium, the transition rates satisfy the detailed balance condition: $\gamma_{e\rightarrow g}^j/\gamma_{g\rightarrow e}^j=P_{-1/2}^{j}/P_{1/2}^{j}=\exp(\beta_\mathrm{h}\omega_j)$.
As $N\rightarrow \infty$, the segmented process approaches the ideal isothermal limit.

During the control stroke $C\rightarrow C^{'}\rightarrow D{'}$, the measurement-based control  employs  two composite light fields: 
\begin{eqnarray}
E_g(t)&=&\epsilon\cos(\Delta\omega_{21}^{C}t)+(1-\epsilon)\cos(\Delta\omega_{13}^{C}t)\nonumber\\
    E_e(t)&=&(1-\epsilon)\cos(\Delta\omega_{11}^{C}t)+\epsilon\cos(\Delta\omega_{23}^{C}t).
\end{eqnarray}
 The frequency differences
$\Delta\omega_{11}^{C},  \Delta\omega_{21}^{C}$, $\Delta\omega_{13}^{C}$ and $\Delta\omega_{23}^{C}$ correspond to the energy differences between the following electronic levels:  $S_{1/2}$ and $D_{-3/2}$, $S_{-1/2}$ and $D_{3/2}$,
 $S_{-1/2}$ and $P_{-3/2}$, and $S_{1/2}$ and $P_{3/2}$, respectively. During the stroke $C\rightarrow C^{'}$, the composite light fields $E_g(t)$ or $E_e(t)$  is used to measure whether the electron is in the ground or excited state. 
This field induces transitions between multiple electronic levels, enabling state detection through fluorescence measurements. The control protocol branches into four cases based on the detected photon frequency: (a) Detection of fluorescence at $\Delta\omega_{21}^{C}$: This indicates an incorrect measurement of the ground state ($y=e$). A laser with frequency $\Delta\omega_{21}^{C}$ is applied to excite the electron from $S_{-1/2}$ to $D_{3/2}$,  followed by another laser with frequency $\Delta\omega_{22}^{C}$  to excite the electron from 
$D_{3/2}$ to $P_{3/2}$.  The electron then spontaneously decays to $S_{1/2}$, achieving a energy level inversion.  This inversion operation takes time $\tau_c$.  The system is then driven via the external field $\omega^\mathrm{hc}_\mathrm{s}(t)=\omega_{C}(1-t/\tau_\mathrm{hc})+\omega_{D^{'}}t/\tau_\mathrm{hc}$  to perform an adiabatic expansion from $\omega_{C}$ to $\omega_{D^{'}}$. Subsequently, the system is modulated back to its initial state, using the same procedure as the isothermal expansion stroke. (b)  Detection of fluorescence at $\Delta\omega_{13}^{C}$: This indicates a correct measurement of the ground state ($y=g$). After waiting for a time $\tau_c$, the system undergoes an  adiabatic expansion and an   isothermal compression  same as case (a). (c) Detection of fluorescence at $\Delta\omega_{11}$: This indicates an correct measurement of the excited state ($y=e$). A laser with frequency $\Delta\omega_{11}^{C}$ is applied to excite the electron from 
$S_{1/2}$ to $D_{-3/2}$, followed by another laser with frequency $\Delta\omega_{12}^{C}$ to excite the electron from $D_{-3/2}$ to $P_{-3/2}$. The electron then spontaneously decays to $S_{-1/2}$, achieving a energy level inversion. Then we  perform the same operation as case (a) to the system. (d) Detection of fluorescence at $\Delta\omega_{23}^{C}$: This indicates a incorrect measurement of the excited state ($y=g$). After waiting for a time $\tau_c$, the system undergoes an  adiabatic expansion and an   isothermal compression same as case (a).

To obtain the average values of work and heat, multiple process cycles are required. During each cycle, the fluorescence frequency, the change in the population of harmonic oscillator during the isothermal compression process, and the population changes after completing one cycle are recorded.

Next, the occurrences of each fluorescence frequency are counted and divided by the total number of cycles to determine the probability of each fluorescence. The corresponding heat absorbed from the cold and work for each fluorescence are then summed and divided by its occurrence count. This operation is used to calculate the average  heat and work  associated with that specific fluorescence.

Finally, the probability of each fluorescence occurrence is multiplied by its corresponding work and heat absorbed from the cold reservoir to obtain the average work and heat of the CIE after one complete cycle. Finally, the CIE efficiency can be determined by the equation: $\eta_\mathrm{d}=W_\mathrm{tot}^\mathrm{d}/(W_\mathrm{tot}^\mathrm{d}-Q_\mathrm{c}^\mathrm{d})$.
\end{widetext}


\begin{thebibliography}{99}
\bibitem{sup1}At point $C^{'}$, the system state $x_{C^{'}}$ is determined by the system state $x_C$ and the measurement outcome $y$.  The post-control state 
$x_{C^{'}}$ can be expressed as a function of the  state $x_{C}$ and measurement outcome 
$y$ following the relations: $x_{C^{'}}(x_C=e,y=e)=g$, $x_{C^{'}}(x_C=e,y=g)=e$, $x_{C^{'}}(x_C=g,y=e)=e$, $x_{C^{'}}(x_C=g,y=g)=g$. In the cases $x_{C^{'}}(x_C=e,y=e)=g$ and $x_{C^{'}}(x_C=g,y=e)=e$ the demon injects work $\varepsilon_C^g-\varepsilon_C^e=-\omega_C(n_e-n_g)$ and $\varepsilon_C^e-\varepsilon_C^g=\omega_C(n_e-n_g)$ into the system. However, in the cases $x_{C^{'}}(x_C=e,y=g)=e$ and $x_{C^{'}}(x_C=g,y=g)=g$, the demon performs zero work.
\end{thebibliography}
\end{document}